\pdfoutput=1
\documentclass{JINST}
\usepackage{graphicx}
\usepackage{amssymb}
\usepackage{makeidx}
\usepackage{amsfonts}
\usepackage{amstext}
\usepackage{amsmath}
\usepackage{amsbsy}
\usepackage{wasysym}
\usepackage{fontenc}
\usepackage{times}
\usepackage{mathptmx}

\title{Experimental observation of an extremely high electron lifetime with the ICARUS-T600 LAr-TPC}

\author{
M.~Antonello$^a$,
B.~Baibussinov$^b$,
P.~Benetti$^c$,
F.~Boffelli$^c$, 
A.~Bubak$^k$,
E.~Calligarich$^c$,
S.~Centro$^b$,
A.~Cesana$^d$,
K.~Cie\'slik$^e$,
D.B.~Cline$^f$,
A.G.~Cocco$^g$,
A.~Dabrowska$^e$,
A.~Dermenev$^h$,
R.~Dolfini$^c$\thanks{Deceased},
A.~Falcone$^c$, 
C.~Farnese$^b$\thanks{Corresponding author},
A.~Fava$^b$,
A.~Ferrari$^i$,
G.~Fiorillo$^g$,
D.~Gibin$^b$,
S.~Gninenko$^h$,
A.~Guglielmi$^b$,
M.~Haranczyk$^e$,
J.~Holeczek$^k$,
M.~Kirsanov$^h$,
J.~Kisiel$^k$,
I.~Kochanek$^k$,
J.~Lagoda$^j$,
S.~Mania$^k$,
A.~Menegolli$^c$,
G.~Meng$^b$,
C.~Montanari$^c$,
S.~Otwinowski$^f$,
P.~Picchi$^l$,
F.~Pietropaolo$^b$,
P.~Plonski$^m$,
A.~Rappoldi$^c$,
G.~L.~Raselli$^c$,
M.~Rossella$^c$,
C.~Rubbia$^{a,i,o}$,
P.~R.~Sala$^d$,
A.~Scaramelli$^d$,
E.~Segreto$^a$,
F.~Sergiampietri$^n$,
D.~Stefan$^d$,
R.~Sulej$^{j}$,
M.~Szarska$^e$,
M.~Terrani$^d$,
M.~Torti$^c$,
F.~Varanini$^b$,
S.~Ventura$^b$,
C.~Vignoli$^a$,
H.G.~Wang$^f$,
X.~Yang$^f$,
A.~Zalewska$^e$,
A.~Zani$^c$,
K.~Zaremba$^m$\\
\llap{$^a$}INFN - Laboratori Nazionali del Gran Sasso, Assergi, Italy\\
\llap{$^b$}Dipartimento di Fisica e Astronomia  Universit\`a di Padova e INFN, Padova, Italy\\
\llap{$^c$}Dipartimento di Fisica Universit\`a di Pavia e INFN, Pavia, Italy\\
\llap{$^d$}INFN, Milano, Italy\\
\llap{$^e$}H.Niewodnicza\'nski Institute of Nuclear Physics, Krak\'ow, Poland\\
\llap{$^f$}Department of Physics, UCLA, Los Angeles, USA\\
\llap{$^g$}Dipartimento di Scienze Fisiche Universit\`a Federico II di Napoli e INFN, Napoli\\
\llap{$^h$}Institute for Nuclear Research of the Russian Academy of Sciences, Moscow, Russia\\
\llap{$^i$}CERN, Geneva, Switzerland\\
\llap{$^j$}National Centre for Nuclear Research, Otwock/Swierk, Poland\\
\llap{$^k$}Institute of Physics, University of Silesia, Katowice, Poland\\
\llap{$^l$}INFN Laboratori Nazionali di Frascati, Frascati, Italy\\
\llap{$^o$}GSSI, L'Aquila, Italy\\
\llap{$^m$}Institute for Radioelectronics, Warsaw Univ. of Technology, Warsaw, Poland\\
\llap{$^n$}INFN, Pisa, Italy\\
E-mail: \email{christian.farnese@pd.infn.it}
}

\abstract{The ICARUS T600 detector, the largest liquid Argon Time  Projection Chamber (LAr-TPC) 
realized after many years of R$\&$D  activities, was installed and successfully operated for 3 years at  
the INFN Gran Sasso underground Laboratory. One of the most  important issues was the need of an 
extremely low residual  electronegative impurity content in the liquid Argon, in order to  transport the 
free electrons created by ionizing particles with very small attenuation along the drift path. 
The solutions adopted  for the Argon recirculation and purification systems have permitted  to reach 
impressive results in terms of Argon purity and a free electron lifetime exceeding 15 ms, 
corresponding to about 20 parts per trillion of $O_2$-equivalent contamination, 
a milestone for any future project involving LAr-TPCs and the development of higher detector mass scales. 
}


\begin{document}

\section{Introduction}

The innovative liquid Argon Time Projection Chamber (LAr-TPC)
detection technique, proposed by C. Rubbia~\cite{rubbia77}, observes
ionizing events in neutrino processes or other rare events with a
performance comparable to the one of a traditional bubble chamber.
The LAr-TPC is fully electronic, continuously sensitive and
self-triggering. Operated at atmospheric pressure with a cheap and
abundant cryogenic noble liquid, it offers both visual and
calorimetric determinations of the recorded events. The operating
principle is based on the fact that in highly purified liquid Argon
(or Xenon) free electrons from ionizing particles can be efficiently
transported over macroscopic distances (meters) with the help of a
uniform electric field to a multi-wire anodic structure placed at the
end of the drift path. 

The ICARUS Collaboration has developed the LAr-TPC technology from
prototypal dimensions to the mass of almost 800 tons of liquid Argon
with the so-called T600 detector~\cite{t600}, installed in the underground
INFN-LNGS Gran Sasso Laboratory near Assergi. Its successful and
extended operation has demonstrated the enormous potentials of this
novel detection technique, developing a vast physics program~\cite{t600_jinst, ica_nue1, ica_nue2}
 and the simultaneous observation of neutrinos both from
the CNGS beam at a distance of 730 km from CERN and from cosmic rays.

As a fundamental requirement for the LAr-TPC performance,
electronegative impurities (main-ly O$_2$, H$_2$O and CO$_2$) must be kept at
extremely low concentrations. This goal can be achieved with a
continuous LAr recirculation system with UHV construction techniques
and dedicated filters. For instance, a $\sim$ 5 m free electron attenuation
length, equivalent to 3 ms lifetime  in an electric guide field of 500
V/cm requires the O$_2$-equivalent concentration as low as 
100 parts per trillion (ppt). 

Laboratory studies initiated around 1987 with remarkably long free
electron lifetimes were followed by a gradual industrialization of the
technique. Already in 2001, only 2 months after the beginning of the
first surface test run of one half of the T600 (275 m$^3$)~\cite{t600}, an
electron lifetime of 1.8 ms was observed.  During 2009, the value of
21 ms was measured at the Laboratori Nazionali di Legnaro (INFN-LNL)
~\cite{icarino}, however on a much smaller scale (120 liters prototype with 38 kg
active mass). More recently, in the three years of T600 underground
operation at the LNGS, electron lifetimes constantly exceeding several
ms were obtained.  During the last part of this data taking, a free
electron lifetime exceeding 15 ms has been achieved. The analysis
method and the most recent results of the measurement of the LAr
purity are here described.

\section{The ICARUS T600 detector}

The ICARUS T600 detector consists of a large cryostat filled with
about 760 tons of ultra-pure liquid Argon and split into two
identical, adjacent modules. A more detailed description can be found
elsewhere~\cite{t600, t600_jinst}. Each module houses two TPCs with 1.5 m maximum
drift path, sharing a common central cathode. A uniform electric field
(E$_{drift}$ = 500 V/cm) drifts ionization electrons with v$_D$ $\sim$ 1.6 mm/$\mu$s 
velocity towards the anode, consisting of three wire arrays and a
stereoscopic event reconstruction. A total of 53248 wires are
deployed, with a 3 mm pitch, oriented on each plane at a different
angle (0$^\circ$, +60$^\circ$, -60$^\circ$) with respect to the horizontal direction.  By
appropriate voltage biasing, the first two wire planes (Induction1 and
Induction2) record signals in a non-destructive way; finally the
ionization charge is collected and measured on the last plane
(Collection). 

The electronics was designed to allow continuous read-out,
digitization and independent waveform recording of signals from each
wire of the TPC. The read-out chain is organized on a 32-channel
modularity. Signals of the charge sensitive front-end amplifiers have
been digitized with 10-bits ADCs with 400 ns sampling channels. The
overall gain is about 1000 electrons for each ADC count, setting the
signal of minimum ionizing particles (m.i.p.) to $\sim$ 15 ADC counts. The
average electronic noise is 1500 electrons, compared with the $\sim$ 15000
free electrons produced by a m.i.p. in 3 mm,  leading to a signal to
noise ratio S/N $\sim$ 10. The gain uniformity has been measured with an
accuracy of about 5\%, determined by the uncertainties on the adopted
calibration capacitances. 

In order to determine the absolute position of the track along the
drift coordinate, the measurement of the absolute time of the ionizing
event provided by a conventional photo-multiplier (PMT) system
detecting the prompt scintillation light in LAr has been combined with
the information coming from the electron drift velocity.

One thermal insulation vessel surrounds the two modules: between the
insulation and the aluminium containers a thermal shield is placed,
with boiling Nitrogen circulating inside to intercept the heat load
and maintain the cryostat bulk temperature uniform (within 1 K) and
stable at 89 K. Nitrogen is stored in two 30 m$^3$ LN2 reservoirs. The
temperature is fixed by the equilibrium pressure in the tanks (2.1
bar, corresponding to about 84 K), which is kept stable in a steady
state by a dedicated re-liquefaction system of twelve cryo-coolers (48
kW global cold power), thus guaranteeing the safe operation in a
closed-loop. 

To keep the electronegative impurities in LAr at a very low
concentration level, each module is equipped with two gasseous Argon (GAr) and
one LAr recirculation/purification 
systems~\cite{t600_jinst, chiara, ica_cryo}. Argon
gas is continuously drawn from the cryostat ceiling and, re-condensed,
drops into Oxysorb$^{TM}$ filters and finally returns to the LAr
containers. LAr instead is recirculated by means of an immersed,
cryogenic pump ($\sim$ 2 m$^3$/h~\cite{t600_jinst}, full volume recirculation in 6 days) and is
purified through standard Hydrosorb$^{TM}$ / Oxysorb$^{TM}$ filters before being
re-injected into the cryostats. LAr is extracted at 1.5 m from the
floor on one side of the vessel, purified and injected back at the
opposite longitudinal side (20 m apart) through several nozzles
uniformly distributed close to the floor of the vessel.  Convective
motions induced by heat losses from the module walls ensure a fast and
almost complete LAr mixing, minimizing the fluctuations of the
relevant parameters, such as LAr density, temperature and purity.
 
\section{Determination of the free electron lifetime in LAr}

The electron lifetime $\tau_{ele}$\footnote{Free electron lifetime is the average capture time of a free ionization electron by an electronegative impurity in LAr. This physical parameter depends in principle on the electric field since above 200 V/cm the drift electrons have more than thermal energy and their cross section for capture can therefore depend on the field. In this paper the free electron lifetime refers to the measurement at the nominal electric field of 500 V/cm in the ICARUS T600.} 
in LAr-TPC has been measured with the help
of the attenuation of the charge signal of traversing cosmic-ray muon
tracks as a function of the electron drift distance.

\begin{figure}[htbp]
\centering
\includegraphics[width=.7\textwidth]{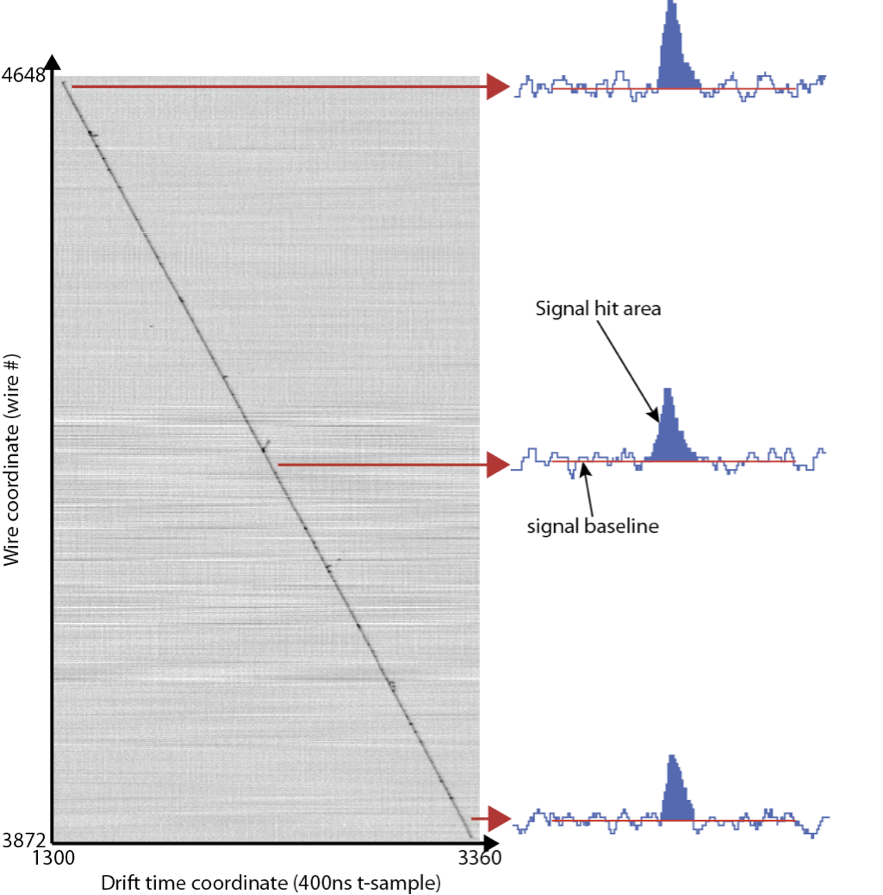}
\caption{Example of a track used for purity measurement extending over 776 wires and 2060 t-samples, corresponding to a drift time of 824 $\mu$s. 
Signals of three different hits are also shown.}
\label{track}
\end{figure}

A new precise method is here introduced to measure the attenuation $\lambda$ =
1/$\tau_{ele}$  of the actual ionization charge signal produced 
by the particle energy deposition in the ICARUS T600 events, as a
function of the drift distance from the wire planes. An automatic
procedure based on the recognition of the track pattern has been used
to provide a first pre-selection of candidates for the purity
measurement. An example of a muon track is shown in Figure~\ref{track}.

The charge signals have been measured in the Collection plane,
removing channels with a r.m.s. noise exceeding 3 ADC counts (3000
electrons, twice the average noise). In order to provide a precise
estimate of the signal attenuation only events with a track with > 100
wires and > 94 cm along the drift coordinate 
(corresponding to a drift time larger than 600 $\mu$s)
have been selected. Tracks with evident
associated electromagnetic showers, with large number of delta-rays,
tracks interrupted by noisy wires and events with multiple tracks have
been rejected. About 20\% of the muon tracks survive these criteria. 

The ionization charge is measured by the area of the signal pulse
above the local baseline level. Single delta rays emitted at large
angle along the track are removed by a recursive linear fit of the
track (wire number versus the drift time), if at more than 3 mm
distance from the track. As the result of these procedures, surviving
tracks have on the average about 200 points. 

A reliable fit of the charge attenuation along each track requires 
Gaussian-like distributions. This can be achieved truncating the asymmetric
Landau tail of the dE/dx depositions, making the distribution more
symmetric around the most probable value. However, to obtain a precise
determination of the charge attenuation it is necessary to apply the
truncation method uniformly along the track.  The procedure is here
elucidated in more detail. 

As a first step the fitting procedure is applied to each track
splitting it into shorter ($\sim$ 10) equal length segments in which the
attenuation is negligible compared to the Landau fluctuations. Inside
each segment the upper and lower 10\% of the signals are discarded to
mitigate anomalous fluctuations in the energy deposition, while
retaining enough statistics. Then the provisional attenuation value is
determined averaging over a track sample for which the LAr purity is
not expected to vary appreciably (typically of the order of 100 tracks
collected during half a day).

\begin{figure}[htbp]
\centering
\includegraphics[width=.6\textwidth]{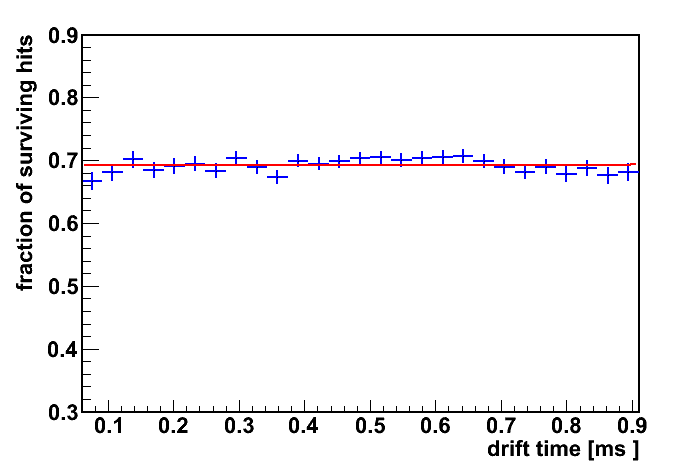}
\caption{Fraction of surviving hits after the truncation method application as a function of the hit drift time.
 A sample of about 500 muon tracks has been studied.The linear fit (continuous line) is fully compatible with a uniform fraction of surviving hits along the drift time (slope: 0.0018 $\pm$ 0.01 ms$^{-1}$).}
\label{trunk_surv}
\end{figure}

\begin{figure}[htbp]
\centering
\includegraphics[width=.8\textwidth]{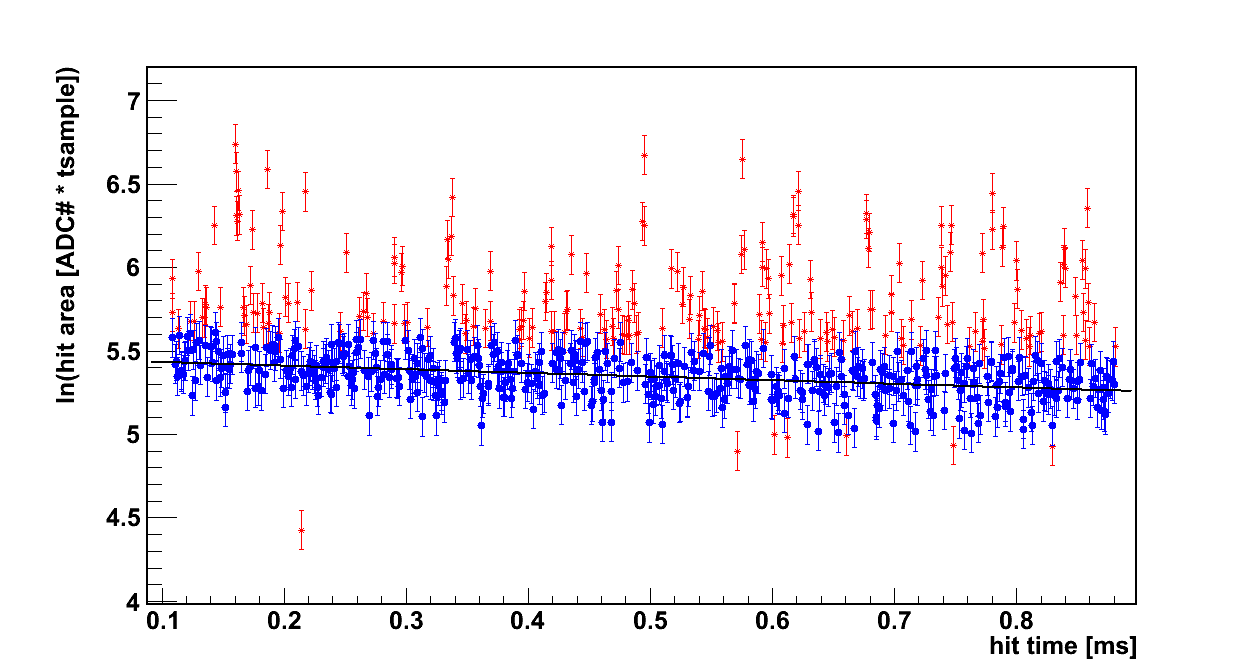}
\caption{Pulse hit area as a function of the drift time for the track shown in Figure 1; 
in red star the $\sim$ 230 hits that are removed by the truncation method, 
in blue circle the $\sim$ 510 surviving hits. The linear fit of the logarithm of the 
hit signal vs. drift time used to extract the electron signal attenuation is also shown 
(black line): for this event $\lambda_T$ = (0.212 $\pm$ 0.022) ms$^{-1}$.}
\label{trunk_fit}
\end{figure}

As a second step and following the attenuation value of the first
iteration, the truncation method is applied to each track without any segmentation,
applying a 30\% upper Landau like cut and excluding the lower 1\% of
the signals. This procedure guarantees a signal distribution centered
around the most probable dE/dx value; its correct application on the
data is demonstrated by the observed uniformity of the truncation
method along the track as shown in Figure~\ref{trunk_surv}. A linear fit of the
logarithm of the survived hit areas versus drift time along the whole
track provides the estimation of attenuation $\lambda_T$ associated to the
selected track (Figure~\ref{trunk_fit}). 

\begin{figure}[htbp]
\centering
\includegraphics[width=.4\textwidth]{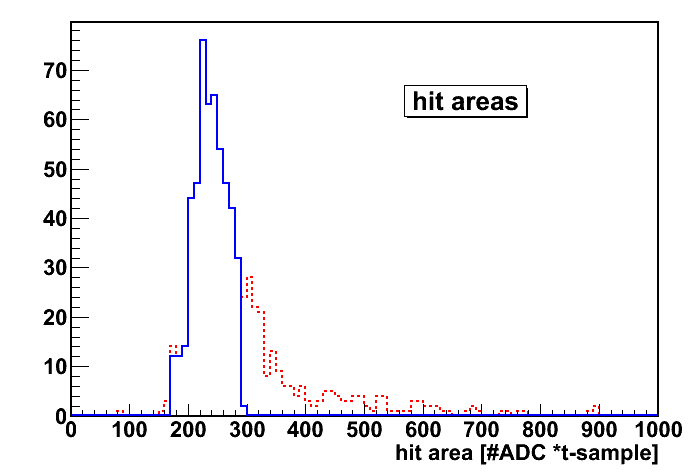}
\caption{Landau distribution of the hit areas corrected by the charge attenuation (dotted line) for the muon track in Figure 1; the effect of the truncation procedure is also shown (continuous line); the r.m.s. of the truncated distribution for this muon track is $\sim$ 12\%.}
\label{landau}
\end{figure}

The resulting distribution of the signals from single wires due to the
above described truncated Landau distribution with the inclusion of
the modest effect of electronic noise has an observed r.m.s. width $\sigma_i$
 $\simeq$ 14\% in agreement with previous estimations~\cite{ica_nue1} (see Figure~\ref{landau}).

\begin{figure}[htbp]
\centering
\includegraphics[width=.6\textwidth]{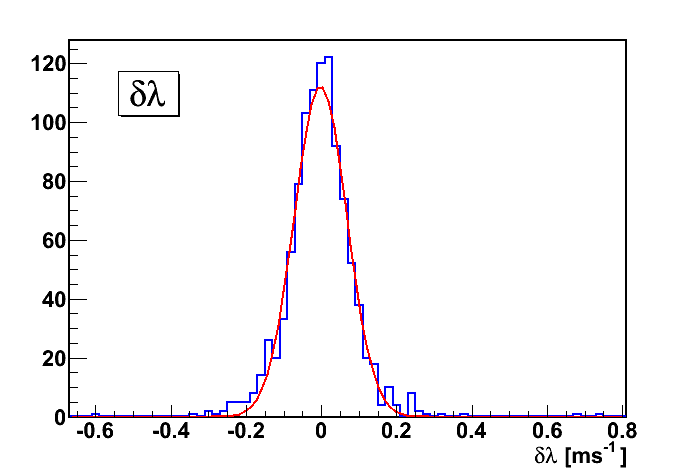}
\caption{Distribution $\delta\lambda_T$ defined as the difference between the single-track  $\lambda_T$ measurements and the corresponding average value. The mean value and the width of the distribution obtained from the gaussian fit are (-0.0029 $\pm$ 0.0022) ms$^{-1}$ and (0.07$\pm$ 0.002) ms$^{-1}$ respectively.}
\label{lambda_res}
\end{figure}

\begin{figure}[htbp]
\centering
\includegraphics[width=.5\textwidth]{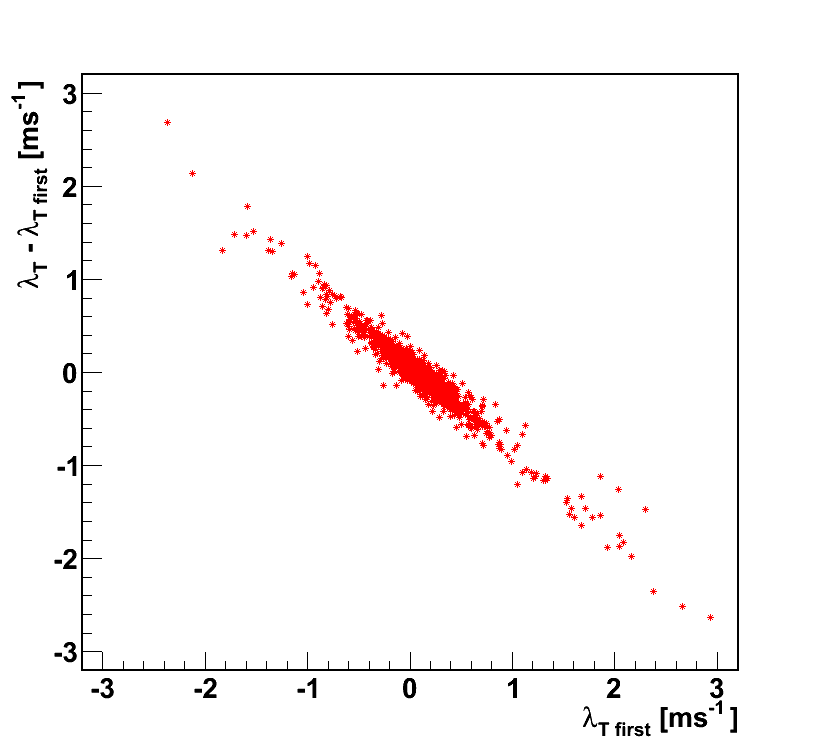}
\caption{Scatter plot of the $\lambda_T$-$\lambda_{T First}$  against $\lambda_{T First}$, where $\lambda_{T First}$ is the measurement obtained for the track in the first step: the visible strong anti-correlation demonstrates that the second step results in an improvement of the $\lambda_T$ measurement, producing a sizable reduction of the observed width of the $\lambda_T$ distribution.}
\label{vs}
\end{figure}

The signals for single wires are then combined into tracks.  As shown
in Figure~\ref{lambda_res}, the observed r.m.s. width $\sigma_\lambda$ of the  $\lambda_T$ distribution is
$\sigma_\lambda$ $\simeq$  0.07 ms$^{-1}$, unaffected by the 
actual $\lambda_T$, as verified experimentally for $\lambda_T$ up to 1 ms$^{-1}$. 
This result is in rough agreement with the naively expected 
value $\sigma_\lambda$ $\simeq$ $\sqrt(12/N)$ $\sigma_i$/$\Delta$T 
$\simeq$ 0.05 ms$^{-1}$, indipendent 
from $\lambda_T$ and obtained combining the r.m.s. noise of the single wires ($\sigma_i$ $\simeq$ 14\%), 
the average number of hits along the track (N $\simeq$ 190) and the
average drift time interval ($\Delta$T $\simeq$  0.70 ms). 
The comparison between the $\lambda_T$ obtained in the first and in the second step 
is also shown in Figure~\ref{vs}: the application of the second step in the method 
allows to obtain a better removal of the Landau tail of the dE/dx distribution and so a 
better fit of the logarithm of the survived hit areas versus drift time, producing a sizable 
reduction of the observed r.m.s. width of the $\lambda_T$ distribution.

As an example for an event sample of $\sim$  100 tracks (corresponding to about 
half a day of data taking), the error on $\lambda$ is 0.007 ms$^{-1}$ allowing to
reach a sensitivity on the electron attenuation length of 0.018 ms$^{-1}$
at 99\% C.L. In terms of free electron lifetime, this implies that
values up to 20 ms can be precisely measured in such a relatively
short time interval of data taking. 

\section{Experimental measurement of the lifetime}

\begin{figure}[htbp]
\centering
\includegraphics[width=.75\textwidth]{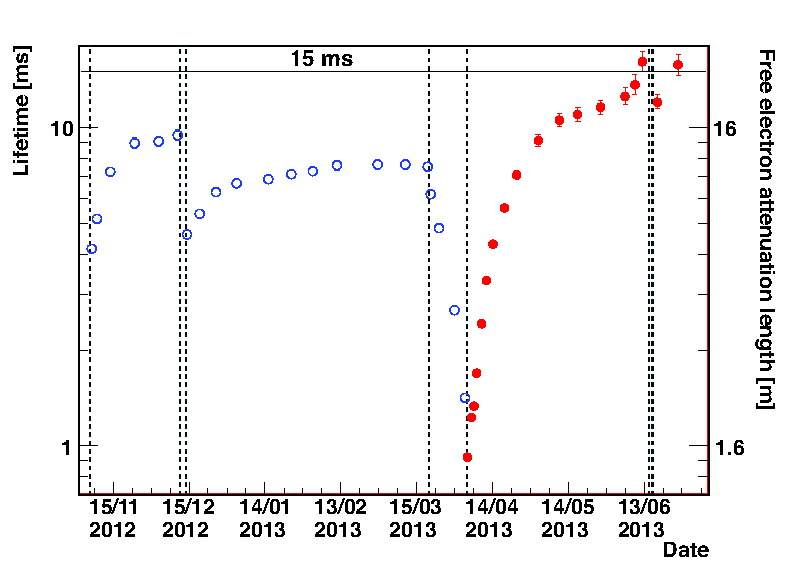}
\caption{Electron lifetime $\tau$ of the East module for the last part of the ICARUS data taking: in red full points the measurements with the new pump with external motor are shown. The dashed vertical lines represent the stops and restart of the LAr recirculation; during this period the GAr recirculation system continued to operate. }
\label{tau_trend}
\end{figure}

The through going cosmic rays collected at the rate of $\sim$ 3100 muons
per day have been used to measure the free electron lifetime in the
ICARUS-T600 providing an almost ideal source of continuous
calibration.  The LAr purity trend in the T600 East module (Figure~\ref{tau_trend})
is here shown for the last few months of operation; each data point
and the related errors are obtained averaging over $\sim$ 100 muon tracks
collected in about half a day.

The analysis of the LAr purity demonstrates that the ICARUS detector
has operated correctly only when both circulation systems are
operational. The interruption of the liquid recirculation system for
pump maintenance resulted in a rapid decrease of the electron lifetime
that was restored promptly as the recirculation system was
reactivated. 

In April 2013 a major upgrade of the LAr recirculation system was
performed in the East cryostat~\cite{chiara, chiara2, ica_cryo}. The ACD CRYO pump used during
the first 2 years of run was replaced with a new Barber Nichols
BNCP-32C-000 with an external motor similar to the ones used in the
LN2 circuit, which worked in a very efficient and reliable way without
frequent stopping. During the 2 week stop of the LAr recirculation,
required for the new pump installation, $\tau_{ele}$ rapidly decreased below 1
ms. After the new pump was switched on, the electron lifetime started
increasing at a rate faster than before (see Figure~\ref{tau_trend}). At the end of
the ICARUS data taking the electron lifetime was still rising and the
last measurement before the detector stop resulted to be 16.1$^{+1.3}_{-1.1}$
ms corresponding to a maximum signal attenuation of 6\% at 1.5 m drift
distance. The remarkable LAr purity obtained in the large T600
detector approaches the result of $\tau_{ele}$ = 21 ms previously obtained
with the smaller LAr-TPC prototype of INFN-LNL~\cite{icarino}.

In view of the very large dimensions of the detector, the uniformity
of the observed lifetime over the volume is a crucial element that has
to be demonstrated. The level of accuracy achieved for the charge
attenuation measurements along single muon tracks allows estimating
the uniformity of the LAr purity in different regions along the 20 m
detector length. To achieve uniformity, as already pointed out, in
each cryostat the injection of the LAr is located at one side $\sim$ 2 m
below the surface and the extraction at 20 m apart at the opposite
side. 

\begin{figure}[htbp]
\centering
\includegraphics[width=.85\textwidth]{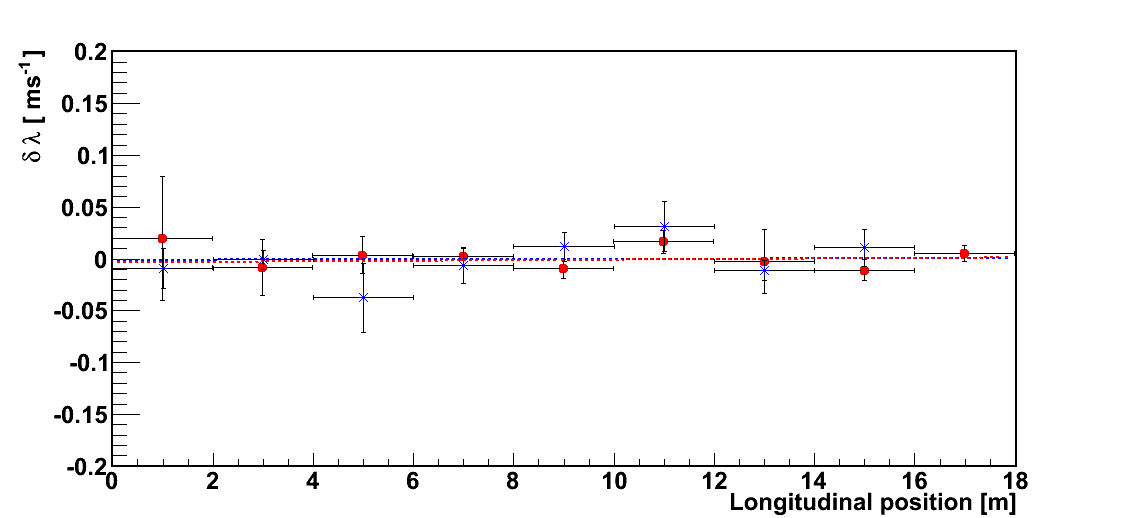}
\caption{The measured variation of the level of impurities in the East cryostat along the longitudinal direction. Red circles refer to the left chamber, blu stars to the right one. The dashed lines are the linear fits in both chambers. The fit results are amply compatible with a uniform LAr purity across the length of the whole detector: slope for the left chamber (8.8 $\pm$ 90) $\cdot$ 10$^{-5}$ ms$^{-1}$m$^{-1}$, slope for the right chamber (2.7 $\pm$ 11) $\cdot$ 10$^{-4}$ ms$^{-1}$m$^{-1}$.}
\label{uniformity}
\end{figure}

In order to investigate the uniformity of the lifetime along the detector, 
a sample of 1000 almost vertical cosmic muon tracks
in the East cryostat have been selected in different periods over the
last 2 months of the ICARUS data taking in which the signal
attenuation was 0.06 < $\lambda$ < 0.10 ms$^{-1}$. The tracks have been
automatically reconstructed in 3D in order to define their position
along the detector longitudinal direction, which has been divided in 9
regions of 2 m each both for the left and the right chamber.  For each
region, the distribution of the $\delta\lambda$ parameter, defined as the
difference between the $\lambda_T$ associated to the track and the $\lambda$ value
measured in the considered period, has been determined and the mean
value for the tracks inside the selected 2 m region has been
extracted. The trend of the $\delta\lambda$ parameter as measured along the
longitudinal direction of the cryostat (Figure~\ref{uniformity}) shows a complete
uniformity of the LAr impurities: indeed the linear fit of $\delta\lambda$ presents
a slope compatible with zero within one sigma in both chambers. A
similar behaviour has been observed also in a period during which the
LAr purity was rapidly increasing from 3.3 ms to 5.6 ms. These results
are in agreement with the assumption that convective motions in the
T600 detector are sufficient to mix LAr faster than the typical
recirculation time (6 days), thus minimizing any fluctuations of the
electron lifetime within the detector volume.

\section{Independent verifications of the results}

\subsection{Muons from neutrino events}

The LAr purity measurement method was validated applying the
attenuation $\lambda$ measured with the cosmic muons to an independent sample
of muon tracks from CNGS neutrino interactions in the upstream rock
collected in the same period of time. In the considered data sample
the electron lifetime was measured to be between 8.8 and 9.5 ms, quite
stable within the measurement errors.

The 254 selected CNGS muons, entering the T600 module and travelling
almost parallel to the wire planes, have been automatically
reconstructed in 3D~\cite{ica_3drec} and the dE/dx associated to each hit along the
track has been estimated as well as the related position along the
drift. The drift path has been split into 15 bins of 10 cm each and
the last bin has been discarded due to small electric fields
distortions in the cathode region.  For each event, the dE/dx
distributions are constructed for every drift bin, requiring at least
30 entries per bin to extract the most probable dE/dx value by fitting
the charge signal distribution with the convolution of Landau and
Gaussian functions. The most probable value of the energy loss has
been preferred to the average energy loss because of its negligible
dependence on muon momentum in the CNGS energy range. The available
event statistics allows achieving in all the 15 bins a statistical
precision better than 1\%. 

\begin{figure}[htbp]
\centering
\includegraphics[width=.8\textwidth]{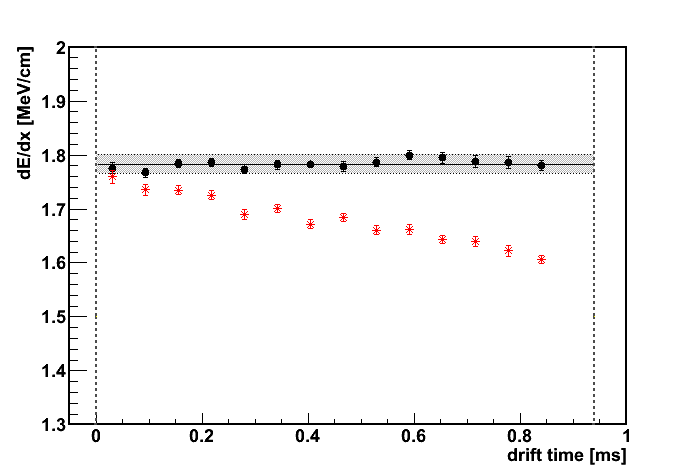}
\caption{The average of the reconstructed most probable value of dE/dx for a sample of CNGS muons is shown as a function of the drift coordinate before (red stars) and after (black points) the correction for the measured $\lambda$ = 1/$\tau_{ele}$ = 0.11 ms$^{-1}$. Only the statistical errors are given. The residual 1\% systematic uncertainty band is also shown to guide the eye.}
\label{cngs_mu}
\end{figure}

As a result, the average dE/dx of the analyzed CNGS tracks, corrected
by the $\lambda$ measurement, is independent from the drift coordinate within
the errors, demonstrating the reliability of the LAr purity estimation
method (Figure~\ref{cngs_mu}).

 \subsection{Monte Carlo events}

The above described method for the purity measurement has been tested
on a 9000 muon track sample in the T600 LAr-TPC generated with the
energy spectrum and the angular distribution of cosmic rays measured
at LNGS by a dedicated Monte Carlo program based on the FLUKA code~\cite{fluka1, fluka2}.
 The simulation includes the free electron longitudinal
diffusion, the electronic response and the noise measured in the
detector as well as the signal attenuation due to the impurities in
LAr ~\cite{t600}. The method has been tested for electron lifetime in the 1.5 - 20
ms interval focusing in particular on the cases 
in which the expected attenuation is relatively small (e.g. a
$\sim$ 10\% signal attenuation for the maximum drift length is expected for
$\tau_{ele}$ $\sim$ 10 ms). According to the described selection criteria, about
2000 simulated tracks were retained as candidates for LAr purity
measurements.

\begin{figure}[htbp]
\centering
\includegraphics[width=.6\textwidth]{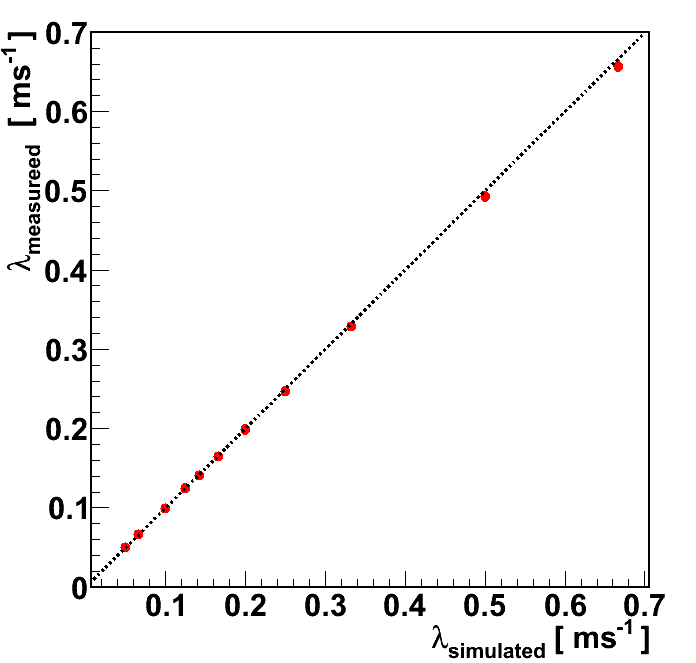}
\caption{The signal attenuation $\lambda$ as measured with the Monte Carlo events as a function of the simulated attenuation values. A small difference is visible, however well within the measurement errors. The interval of $\lambda$ corresponds to an electron lifetime between 1.5 and 20 ms.}
\label{calibr}
\end{figure}

The observed r.m.s. width $\sigma_\lambda$ of the 
electron attenuation $\lambda_T$ distribution 
in the MC events is $\sigma_\lambda$ $\simeq$  0.06 ms$^{-1}$,
in a reasonable agreement with experimental results (Figure~\ref{lambda_res}).  
The results of the presented method applied to simulated data at different
purity values, summarized in Figure~\ref{calibr}, show the measurement
reliability in a large range of LAr purity values. The residual < 1\%
underestimation of $\lambda$ at LAr high purity values ($\tau_{ele}$  > 3 ms)
corresponds to < 0.3\% effect on the charge signal corrected for the
attenuation effect even for the maximum drift time of 1 ms. For low
purity values ($\tau_{ele}$ < 3 ms) a $\sim$ 1.5\% underestimation is detected
affecting the charge signal correction less than 1\%.

\section{Conclusions}

The successful ICARUS-T600 operation at LNGS with the simultaneous
exposure to both CNGS neutrino beam and cosmic rays, demonstrates the
enormous potential of this detection technique. The ICARUS cryogenic 
system and the solutions adopted
for the argon re-circulation and purification systems permitted to
reach an impressive result in terms of argon purity, which is one of
the key issues for the superb detector performance. A corresponding
free electron lifetime exceeding 15 ms has been obtained corresponding 
to a mean attenuation length of 25 meters, a milestone
for any future project involving liquid argon TPC.

In addition this result demonstrates the effectiveness of the single
phase LAr-TPC detectors paving the way to the construction of huge
detectors with longer drift distances: for example, with the achieved
purity level, at 5 m from the wire planes the maximum signal
attenuation is only $\sim$ 23\%.

\acknowledgments

The results reported in this paper could not have been achieved
without the effort of the LNGS Research and Technical Divisions, and
in particular the Experiment Support Service and the LNGS cryogenic
group, and the Technical Service of INFN Pavia that contributed to the
implementation and operation of the ICARUS T600 cryogenic plant. The
ICARUS Collaboration recognizes the fundamental involvement of the
industrial companies Air Liquide, Stirling Cryogenics BV and Luca
Scarcia, in the realization, operation and maintenance of the
cryogenic plant. The Polish groups acknowledge the support of the
National Science Center: Harmonia (2012/04/M/ST2/00775) funding
scheme.


\begin{thebibliography}{9}

\bibitem{rubbia77} 
C. Rubbia, \emph{The liquid-argon time projection chamber: a new
  concept for neutrino detector}, CERN-EP/77-08 (1977).

\bibitem{t600}
S.Amerio et al, ICARUS Collaboration, \emph{Design, construction and
  tests of the ICARUS T600 detector}, Nucl. Instr. Meth. A527 (2004) 329-410.

\bibitem{t600_jinst}
C. Rubbia et al., ICARUS Collaboration, \emph{Underground operation of
    the ICARUS T600 LAr-TPC: first results}, \jinst{6}{2011}{P07011}.

\bibitem{ica_nue1}
M. Antonello et al., ICARUS Collaboration, \emph{Experimental search 
  for the "LSND anomaly", with the
  ICARUS detector in the CNGS beam}, Eur. Phys. J. C, 73:2345 (2013).

\bibitem{ica_nue2}
M. Antonello et al., ICARUS Collaboration, \emph{Search for anomalies
  in the $\nu_e$ appearance from a $\nu_\mu$ beam}, Eur. Phys. J. C, 73:2599 (2013).

\bibitem{icarino} 
B. Baibussinov et al., \emph{Free electron lifetime achievements in
  liquid Argon imaging TPC}, \jinst{5}{2010}{P03005}.

\bibitem{chiara}
C. Vignoli, \emph{The ICARUS T600 liquid argon purification system},
presented at the 25th International Cryogenic Engineering Conference
and the International Cryogenic Materials Conference in 2014, ICEC
25-ICMC (2014), in publication.

\bibitem{chiara2}
C. Vignoli, \emph{The ICARUS T600 Liquid Argon Detector Operation in the Underground Gran Sasso Laboratory},
proceeding for the 13th Cryogenics 2014, IIR International Conference, 
April 7 - 11, 2014, Prague, Czech Republic.

\bibitem{ica_cryo}
M. Antonello et al., ICARUS Collaboration, \emph{Operation and
  performance of the ICARUS-T600 Cryogenics Plant at The Gran Sasso
  Underground Laboratory}, to be submitted to JINST.

\bibitem{ica_3drec}
M. Antonello et al., ICARUS Collaboration, \emph{Precise 3D track
  reconstruction algorithm for the ICARUS T600  liquid argon time
  projection chamber detector}, Advances in High  Energy Physics,
Volume 2013 (2013), Article ID 260820.

\bibitem{fluka1}
A. Ferrari et al., \emph{FLUKA: a multi-particle transport code},
CERN-2005-10, INFN/TC-05/11 (2005).

\bibitem{fluka2}
T. T. B\"ohlen et al., \emph{The FLUKA code: Developments and
Challenges for High Energy and Medical Applications}, Nuclear Data
Sheets, Volume 120, June 2014, Pages 211-214.

\end{thebibliography}
\end{document}